\documentclass[preprint,superscriptaddress,amsfonts,amssymb,amsmath,showpacs]{revtex4}

\usepackage{amsmath}
\usepackage{epsfig}
\usepackage{graphics}
\usepackage{color}
\usepackage{graphicx}
\usepackage[font={footnotesize,it}]{caption}
\usepackage[colorlinks=true,
            linkcolor=red,
            urlcolor=blue,
            citecolor=green]{hyperref}

\begin{document}

\title{Semiclassical gravitational effects  on the gravitational lensing in the spacetime of topological defects}
\author{Kimet Jusufi}
\email{kimet.jusufi@unite.edu.mk}
\affiliation{Physics Department, State University of Tetovo, Ilinden Street nn, 1200,
Tetovo, Macedonia}
\affiliation{Institute of Physics, Faculty of Natural Sciences and Mathematics, Ss. Cyril and Methodius University, Arhimedova 3, 1000 Skopje, Macedonia}

\author{Farook Rahaman}
 \email{rahaman@associates.iucaa.in}
\affiliation{Department of Mathematics, Jadavpur University,
 Kolkata-700032, India.}

\author{Ayan Banerjee}
 \email{ayan_7575@yahoo.co.in}
\affiliation {Department of Mathematics, Jadavpur University,
 Kolkata-700032, India.}
\affiliation {Astrophysics and Cosmology Research Unit, University of KwaZulu Natal, Private Bag X54001, Durban 4000, South Africa.}
\date{\today }

\begin{abstract}
The theory of gravitational lensing has revealed many generic and fundamental properties of compact objects like black holes and wormholes. In this article, we utilize a recent formulation to compute the quantum effects on the deflection angle of a light ray, namely, the Gauss-Bonnet theorem (GBT) to explore the semiclassical gravitational effects in the spacetime of a point-like global monopole and a cosmic string. Previously, the Gauss-Bonnet theorem [Class. Quant. Grav. 25, 235009 (2008)] was proposed as an alternative way to compute the deflection angle of light in a static, spherically symmetric and asymptotically flat spacetime. In the present article we have used the celebrated GBT that applied to the optical metric as well as the geodesic method in computing the deflection angle. Interestingly one can  observe that we have found an exact result between GBT and the standard approach up to the third-order contributions terms by modifying the domain of integration for cosmic string and global monopole deflection angles. Finally we have considered the time delay in the cosmic string/global monopole spacetime and found that the delay in time is proportional to the linear mass density of the  cosmic string and global monopole parameter, respectively.
\end{abstract}

\maketitle

\section{Introduction}
Over the past few years, there has been growing interest in studying of quantum theory of gravity, but after more than seventy years since, a consistent and complete theory of quantum gravity still seems faraway from us. In recent past, it has been proposed that  phase transitions in the early Universe could lead to the formation of topological defects like domain walls, cosmic strings, textures and monopoles \cite{kibble,vilenkin1,Rahaman2,got,mello}. Cosmologists show  interest in defects as one of the possible sources for the density perturbations which engrained galaxy formation \cite{kunz}. Among the topological defects, the cosmic string plays a significant role  in the large scale structure formation of the Universe. Also  it is proficient of producing observational effects in many areas of physics.
The topological defects has an extensive applicability, like the effects of global monopoles and cosmic strings in the Hawking radiation  \cite{vieira1,vieira2,kimet0}, and furthermore the Lorentz symmetry breaking effects in the cosmic string spacetime has been investigated in Refs. \cite{belich1,bakke1,kimet4}, Landau quantization in the cosmic string spacetime \cite{cunha1}, dependence of the black body force on the nature of topology \cite{cunha2}, scattering and bound states of different spin particles in cosmic string spacetime \cite{andreade1,andreade2}, including the Aharonov-Bohm effect \cite{andreade3}, cosmological constant from  topological defects \cite{R4}, dark matter candidate \cite{R5} etc. 

Cosmic strings plays an important roles in various astrophysical and cosmological studies like gravitational lenses, galaxy formation etc. and created by grand unified phase transitions in the early universe. Barriola and Vilenkin \cite{vilenkin3} showed that vacuum stress-energy may be determined, up to a single dimensionless numerical constant without producing a full renormalization calculation. The obtained vacuum stress-energy tensor then used as a source in the semiclassical Einstein equations for solving the quantum perturbations (in first order of $\hslash$ ) of the metric.

It is known  that  the trace of the energy  stress tensor vanishes,  if the Lagrangian is conformally invariant.  However, in quantized theory it attain a trace  during renormalization. This trace anomaly is a geometrical scalar  comprising with  the metric tensor. 
One can find the trace of the vacuum stress energy for a 
conformally coupled massless free field  as \cite{Hiscock1,Hiscock2,Linet,Rahaman1}

\[{T_\mu}^\mu  = \frac{1}{28802 \pi^2} \left[ a \,C_{\alpha \beta \gamma \delta}C^{\alpha \beta \gamma \delta} + b \left( R_{\sigma \tau}R^{\sigma \tau}-\frac{1}{3} R^2\right) + c ~\Box ~R + d R^2 \right],\]
where the constants $ a, b, c $ and $d$ are coming from the conformal scalar field, which are determined by the spin of the field under
consideration. The remaining symbols have the same meaning as in Riemannian geometry. Hishcock \cite{Hiscock2} has considered these non zero vacuum expectation value of the stress energy tensors as a 
 source to study the semi classical gravitational effects near    
cosmic strings. Similar to cosmic string, the monopole is one of the topological defects,  that can be formed  when the vacuum manifold comprises surfaces, which can’t be contracted continuously to a point.  The monopoles attract  both particle physicists and cosmologists  as it is anticipated to exist in GUT   \cite{vs}  . These monopoles contain Goldstone fields  and have  energy density  inversely proportional to the  distance as $r^{-2}$.  The monopoles have some interesting properties: they exerts no gravitational force on its surrounding  matter \cite{Rahaman3}  however a solid deficit angle is formed around it.  In a pioneering work, Aryal \textit{et al.} derived a Schwarzschild  metric pierced by a static cosmic string \cite{vilenkin2}. Furthermore, Barriola and Vilenkin (BV) discovered  the survival of such a monopole solution in a Schwarzschild background which is resulting from the breaking of global $SO(3)$ symmetry of a triplet scalar field \cite{vilenkin3}. 
Hiscock \cite{Hiscock1} obtained the form of entire vacuum stress energy tensors which are given by 

\[  <{T_\mu}^\nu> = \hbar r^{-4} \left[– D , – ( C + D ) , ( C + D ) , ( C + D ) \right],\] 
 where the constants  $C$ and $D$ are  connected to vacuum expectation value and trace anomaly of the energy momentum tensors. This form is valid for any combination of massless free conformal fields.

Like cosmic string in semiclassical gravity, Hiscock considered the vacuum expectation value of the stress energy tensor in the space-time of a static monopole to obtain semi classical gravitational effects for the quantum perturbations of the metric \cite{Hiscock2}. Inspired by the work, we aim to consider the semiclassical gravitational effects on the deflection of light in the spacetime background with topological defects.

General Relativity (GR) predict that in the presence of a massive body the light is deflected and for experimental confirmation was first provided by the gravitational bending of light by the Sun. This process is called as \textit{gravitational lensing} (GL) and this deflection of light was first discovered (the quasar QSO 0957+561) by Walsh \textit{et al} in 1979 \cite{Walsh}. By contrast, deflections of the light from the source provides direct information concerning distribution of mass independently of its state and nature. The study of light bending when calculated based on the first order expansion of the smallest deflection angle, it is then called weak gravitational lensing (WGL). A number of applications on WGL came through observational cosmology and technical advances in \cite{Klimov,Liebes,Refsdal,Bourassa,RBourassa}. Their quantitative analysis provides specific distribution and evolution of matter. Mellier \cite{Mellier} had studied  most of these recent works regarding on WGL and discuss their impact for cosmology. This led some authors to consider the WGL for studying black holes \cite{Majumdar,Horvath,Gyulchev}. The GL has also of theoretical importance, from a strong field perspective. For instance, Ellis and Virbhadra \cite{KVirbhadra} obtained the lens equation for the Schwarzschild black hole with an asymptotically flat metric \cite{Virbhadra} and more comprehensively by Virbhadra . This success leads to further investigate in studying astrophysical objects like black holes, wormholes, naked singularities, and some other exotic objects in \cite{Perlick,Hasse,Nandi, Eiroa,EEiroa,Clement}. The time delay is another famous classical test experiment confirming Einstein's theory of relativity \cite{weinberg}. It will be interesting to consider this problem in the spacetime of topological defects.

Therefore, it is interesting to study the GL effect caused by the formation of topological defects in the weak gravitational field and a new approach was initiated by Gibbons-Werner \cite{gibbons1} to calculate deflection angle which emphasizes global properties. As a physical application, this method has been applied by several authors such as Werner \cite{Werner} to stationary black holes and Jusufi and others applied to calculate the quantum improved deflection angle for Schwarzschild black hole, wormholes \cite{kimet1,kimet2,kimet3,kimet5},  the effect of the cosmological constant on the rotating cosmic string \cite{kimet6} and very recently application to the Rindler modified Schwarzschild black hole by Sakalli and Ovgun \cite{aovgun}.
In this paper we have applied the GBT to compute the quantum corrected deflection angle of light in the space-time of topological defects. The structure of this paper is as follows. In Section \textbf{II}, we shall calculate the semiclassical gravitational effects by a cosmic string in the deflection angle of a cosmic string spacetime. In Section \textbf{III}, we extend our results by considering the role played by a magnetic flux cosmic string on the deflection of light.  In Section \textbf{IV}, we consider these effects in the spacetime of a point-like global monopole. In Section \textbf{V}, we shall investigate the time delay in the spacetime background of a cosmic string and a global monopole. In Section \textbf{VI}, we draw conclusions. Throughout this paper, we consider a geometric unit system where $c = G = 1$.

\section{Semiclassical gravitational effects by a cosmic string}
To start with let us consider the Einstein field equations by 
considering the non-zero vacuum expectation values of stress-energy tensor $<T_{\mu\nu}>$,
of a quantum field. In particular, this procedure is known as semiclassical approach
to the quantum theory of gravity:
\begin{equation}
G_{\mu\nu}=8 \pi <T_{\mu\nu}>,
\end{equation}
at a linearized level, where the zeroth-order (in $\hbar$) metric generates 
the first order metric perturbation on the zeroth-order string background 
of the vacuum polarisation $ <T_{\mu\nu}>$,  on the space-time geometry.
Using this we obtain space-time metric to first order in $\hbar$ and geometrical units are used, with $G =c = 1$. 

Here we consider a $(3 + 1)$-dimensional cosmic string metric with a mass per unit length $\mu$, located along the z-axis in cylindrically symmetric coordinate is described by the line element
\begin{equation}\label{cs1}
\mathrm{d}s^2= -\mathrm{d}t^2+\mathrm{d}\rho^2+dz^2+\left(1-4\mu\right)^2\rho^2 \mathrm{d}\varphi^2,
\end{equation}
where $\mu$ is dimensionless quantity represents the mass per unit length.
It is notable that Helliwell and Konkowski \cite{Helliwell} and Linet \cite{Linet} have independently calculated
the values of vacuum expectation of the stress-energy tensor of a conformally coupled massless scalar field. The expression of the energy-momentum tensor for the vacuum expectation value in 
$\left(t, \rho, \varphi, z\right)$ coordinate system of Eq. \eqref{cs1} has been found to be 
\begin{equation}
<{T_{\mu}}^{\nu}> = \frac{\hbar}{ 1440 \pi^2 \rho^4}\left[\frac{1}{(1-4\mu)^4}-1\right]  \text{diag}(1, 1, -3, 1).
\end{equation}
This result enables to obtain vacuum expectation value of the stress-energy tensor. Following the procedure adopted by Hiscock \cite{Hiscock1}, the exterior string metric to first order in $\hbar$, take the form of
\begin{widetext}
\begin{equation}\label{cs2}
\mathrm{d}s^2=\left(1-\frac{4 \pi A \hbar }{\rho^2}\right)\left(-\mathrm{d}t^2+\mathrm{d}z^2\right)+\mathrm{d}\rho^2+\left(1-4\mu\right)^2 \rho^2 \left(1+\frac{16 \pi A \hbar}{\rho^2}\right)\mathrm{d}\varphi^2.
\end{equation}
\end{widetext}
For the purpose of this work, let us now consider the following coordinate transformation $z = r \cos\theta$, $\rho = r \sin\theta $
and without loss of generality set $\theta=\pi/2$, the metric \eqref{cs2} in spherical coordinates takes the form
\begin{equation}\label{metric1}
\mathrm{d}s^2=-\left(1-\frac{4 \pi A \hbar }{r^2}\right)\mathrm{d}t^2+\mathrm{d}r^2+(1-4\mu)^2 r^2 \left(1+\frac{16 \pi A \hbar}{r^2}\right)\mathrm{d}\varphi^2.
\end{equation}

Next, we derive null geodesic equations i.e., $\mathrm{d}s^2$ =0, for the above spacetime metric, which is 
\begin{equation}\label{dt}
\mathrm{d}t^2=\frac{\mathrm{d}r^2}{1-\frac{4 \pi A \hbar }{r^2}}+\left[\frac{(1-4 \mu)^2 r^2 \left(1+\frac{16 \pi A \hbar}{r^2}\right) }{1-\frac{4 \pi A \hbar }{r^2}}\right]\mathrm{d}\varphi^2.
\end{equation}
In some applications it is more appropriate to introduce a radial Regge-Wheeler tortoise type coordinate $r^\star$, with a new function  $f(r^\star)$ as follows:
\begin{eqnarray}\label{r}
\mathrm{d}r^\star &=&\frac{\mathrm{d}r}{\sqrt{1-\frac{4 \pi A \hbar }{r^2}}}, \\ \label{f}
f(r^\star)&=& \frac{(1-4\mu) \,r }{\sqrt{1-\frac{4 \pi A \hbar }{r^2}}}\sqrt{1+\frac{16 \pi A \hbar}{r^2}},
\end{eqnarray}

and the line element of the optical metric reads 
\begin{equation}
\mathrm{d}t^2 \equiv g_{ab}^{op} \mathrm{d}x^a \mathrm{d}x^b={\mathrm{d}r^{\star}}^{2}+{f(r^{\star})}^2 \mathrm{d}\varphi^2.
\end{equation}

At this point, it is very important to obtain the Gaussian optical curvature $K$, which is related to the Riemann tensor and can be expressed as \cite{gibbons1}
\begin{eqnarray}
K&=&-\frac{1}{f(r^{\star })}\frac{\mathrm{d}^{2}f(r^{\star })}{\mathrm{d}{%
r^{\star }}^{2}}\\
&=& -\frac{1}{f(r^{\star })}\left[ \frac{\mathrm{d}r}{\mathrm{d}r^{\star }}%
\frac{\mathrm{d}}{\mathrm{d}r}\left( \frac{\mathrm{d}r}{\mathrm{d}r^{\star }}%
\right) \frac{\mathrm{d}f}{\mathrm{d}r}+\left( \frac{\mathrm{d}r}{\mathrm{d}%
r^{\star }}\right) ^{2}\frac{\mathrm{d}^{2}f}{\mathrm{d}r^{2}}\right].\notag
\end{eqnarray}

Making use of Eqs. \eqref{r} and \eqref{f} the Gaussian optical curvature reads
\begin{equation}
K={\frac {-16384\,{A}^{5}{\pi}^{5}{r}^{12}{\hbar }^{5}+16384\,{A}^{4}{
\pi}^{4}{r}^{14}{\hbar }^{4}-2304\,{A}^{3}{\pi}^{3}{r}^{16}{\hbar }^{3
}-96\,{A}^{2}{\pi}^{2}{r}^{18}{\hbar }^{2}-24\,A\pi\,{r}^{20}\hbar }{
 \left( 4\,\pi\,\hbar \,A-{r}^{2} \right) ^{2}{r}^{16} \left( 16\,\pi
\,\hbar \,A+{r}^{2} \right) ^{2}}}.
\end{equation}

Or, after we consider only the linear term in $\hbar$, we find
\begin{equation}  \label{GC1}    
K \simeq  -{\frac {24\,A\pi\,\hbar }{{r}^{4}}}+\mathcal{O}(\hbar^2,A^2).
\end{equation}

The last equation gives our quantum corrected optical Gaussian curvature which will be used in computing the quantum corrected deflection angle of light in the next section.
\newpage

\subsection{Deflection angle}
\begin{figure}[h!] 
\center
\includegraphics[width=0.49\textwidth]{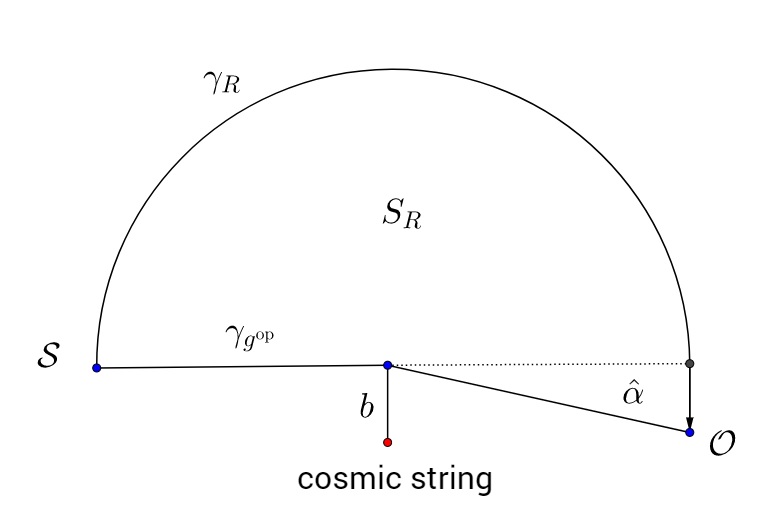} 
\caption{\small \textit{Deflection angle of light in the equatorial plane $(r,\varphi)$ with the deflection angle $\hat{\alpha}$ and the impact parameter $b$. The cosmic string is perpendicular to the $(r,\varphi)$ plane.}}
\label{f1}
\end{figure}

\textbf{Theorem:} \textit{Let  $\mathcal{S}_{R}$ be a non-singular region with boundary $\partial
\mathcal{S}_{R}=\gamma _{g^{op}}\cup \gamma_{R}$, and let $K$ and $\kappa$ be the Gaussian optical curvature and the geodesic curvature, respectively. Then GBT reads}  \cite{gibbons1}
\begin{equation}\label{GBT}
\iint\limits_{\mathcal{S}_{R}}K\,\mathrm{d}A+\oint\limits_{\partial \mathcal{%
S}_{R}}\kappa \,\mathrm{d}t+\sum_{i}\theta _{i}=2\pi \chi (\mathcal{S}_{R}),
\end{equation}
where dA is the element of area of the surface and
 $\theta_{i}$ is the exterior angle at the $i^{th}$ vertex. According to Fig.\,\ref{f1}, we consider a nonsingular domain outside of the light ray, where Euler characteristic for a non-singular domain is  $\chi (\mathcal{S}_{R}) = 1$ and for singular domain $\chi (\mathcal{S}_{R}) = 0$ \cite{gibbons1}. 

In order to find the deflection angle of light, let us first compute the geodesic curvature using the following relation
\begin{equation}
\kappa =g^{\text{op}}\,\left(\nabla _{\dot{%
\gamma}}\dot{\gamma},\ddot{\gamma}\right) ,
\end{equation}
where the velocity and acceleration vectors along the curve $\gamma$ are $\dot{\gamma}$ and $\ddot{\gamma}$, respectively with the unit speed condition $g^{op}(\dot{\gamma},\dot{%
\gamma}) = 1$. Now, if we  allow $R\rightarrow \infty $, then the two jump angles ($\theta _{\mathcal{O}}$, $\theta _{\mathcal{S}}$) become $\pi /2$, or in other words, the sum of jump angles, one for  $\mathcal{S}$ and other for  observer $\mathcal{O}$, holds the following condition $\theta _{\mathit{O}}$ + $ \theta _{\mathit{S}}\rightarrow \pi$ \cite{gibbons1}. 

It follows from the simple geometry that $\kappa (\gamma _{g^{op}})=0$  as $\gamma _{g^{op}}$ is a geodesic. Then we are left to compute the geodesic curvature $\kappa$ with the following relation
\begin{equation}
\kappa (\gamma_{R})=|\nabla _{\dot{\gamma}_{R}}\dot{\gamma}_{R}|,
\end{equation}

 where we have choose $\gamma_{R}:=r(\varphi)=R=\text{constant}$ and $\dot{\gamma}_{R}$ is the velocity vector along the curve ${\gamma}_{R}$. Thus the radial part is evaluated as
\begin{equation}
\left( \nabla _{\dot{\gamma}_{R}}\dot{\gamma}_{R}\right) ^{r}=\dot{\gamma}_{R}^{\varphi
}\,\left( \partial _{\varphi }\dot{\gamma}_{R}^{r}\right) +\tilde{\Gamma} _{\varphi
\varphi }^{r}\left( \dot{\gamma}_{R}^{\varphi }\right) ^{2}, \label{12}
\end{equation}
where  $\tilde{\Gamma}_{\varphi\varphi }^{r}$ denotes the Christoffel symbol associated with the optical metric geometry. Here, the first component of the above expression vanishes, while the second term is being calculated from the unit speed condition $\tilde{g}_{\varphi \varphi}\dot{\gamma}_{R}^{\varphi } \dot{\gamma}_{R}^{\varphi }=1$. Then, we will show that these definitions led to 
\begin{eqnarray}\notag\label{gcurvature}
\lim_{R\rightarrow \infty }\kappa (\gamma_{R}) &=&\lim_{R\rightarrow \infty
}\left\vert \nabla _{\dot{\gamma}_{R}}\dot{\gamma}_{R}\right\vert , \notag \\
&\rightarrow &\frac{1}{R}. 
\end{eqnarray}%

We shall now consider for very large radial distance Eq. \eqref{dt}, yields
\begin{eqnarray}\notag\label{dt1}
\lim_{R\rightarrow \infty } \mathrm{d}t &=&\lim_{R\rightarrow \infty
}\left[\frac{(1-4 \mu) R \left(1+\frac{16 \pi A \hbar}{R^2}\right)^{1/2}}{\left(1-\frac{4 \pi A \hbar }{R^2}\right)^{1/2}}\right] \mathrm{d}\varphi\\
&\to & (1-4 \mu) R \, \mathrm{d}\varphi.  
\end{eqnarray}%

Using relations \eqref{gcurvature} and \eqref{dt1} are automatically leads to the following relation $\kappa (\gamma_{R})\mathrm{d}t = (1-4\mu)\mathrm{d}\,\varphi$. It would be interesting to find the leading orders of the asymptotic deflection angle, we suppose the deflection line as $ r(\varphi) =b/ \sin\varphi$. To achieve the geodesic curvature from GBT it follows
\begin{equation}\label{GBT2}
\iint\limits_{\mathcal{S}_{R}}K\,\mathrm{d}A+\oint\limits_{\gamma_{R}}\kappa \,%
\mathrm{d}t\overset{{R\rightarrow \infty }}{=}\iint\limits_{\mathcal{S}%
_{\infty }}K\,\mathrm{d}A+ (1-4 \mu) \int\limits_{0}^{\pi + \hat{\alpha}}\mathrm{d}\varphi
=\pi,
\end{equation}
where the surface element is given by $\mathrm{d}A=\sqrt{\det g^{op}}\mathrm{d}r^{\star}\mathrm{d}\varphi.$ As we can see from Eq. \eqref{GBT} we need to integrate over the domain $\mathcal{S}_{\infty}$ to find the deflection angle which is quite an amazing result. We already saw that the presence of cosmic string effects the optical geometry. In particular from Eq. \eqref{dt1} we see that our optical metric is not asymptotically Euclidean i.e., $\kappa(\gamma_R) \mathrm{d}t \neq 1$.  One may use a straight line approximation (undeflected light ray) and choose  the light ray as $r=b/\sin\varphi$, however this gives correct result for the deflection angle only to the first order terms.  In our case, as we know, the presence of a cosmic string is encoded via $\varphi \to (1-4\mu ) \varphi$, therefore in contrast to previous works we can make the following choice:
\begin{equation}
r=\frac{b}{\sin\Big[(1-4\mu) \varphi\big]},
\end{equation} 
which of course reduces to the undeflected light ray by taking $\mu\to 0$. Let us substitute this equation into Eq. \eqref{GBT2} then we find that the deflection angle is given in terms of the following integral
\begin{eqnarray}
\hat{\alpha}&=&\frac{4 \pi \mu}{1-4 \mu}-\frac{1}{1-4 \mu}\int\limits_{0}^{\pi}\int\limits_{\frac{b}{\sin \left[(1-4\mu)\varphi \right]}}^{\infty}\left(  -{\frac {24\,A\pi\,\hbar }{{r}^{4}}} \right)\sqrt{\det g^{op}}\,\mathrm{d}r^{\star}\mathrm{d}\varphi.
\end{eqnarray}

The first term can be approximated as 
\begin{equation}
\mathcal{I}_1=\frac{4 \pi \mu}{1-4 \mu}=4 \pi \mu +16 \pi \mu^2+\hdots
\end{equation}

The second term on the other hand gives
\begin{eqnarray}\notag
\mathcal{I}_2 &=&-\frac{1}{1-4 \mu}\int\limits_{0}^{\pi}\int\limits_{\frac{b}{\sin \left[(1-4\mu)\varphi \right]}}^{\infty}\left(  -{\frac {24\,A\pi\,\hbar }{{r}^{4}}} \right)\sqrt{\det g^{op}}\,\mathrm{d}r^{\star}\mathrm{d}\varphi \\ 
&=& \frac{6 \pi^2 \hbar A}{b^2}+\frac{24 \pi^2 \hbar \mu A}{b^2}+\frac{96 \pi^2 \mu^2 \hbar A}{b^2}.
\end{eqnarray}

Thus, we finally find the total deflection angle to be
\begin{equation}\label{a1}
\hat{\alpha}\simeq 4 \pi \mu+ 16 \pi \mu^2+\frac{6 \pi^2 \hbar A}{b^2}+\frac{24 \pi^2 \hbar \mu A}{b^2}+\frac{96 \pi^2 \mu^2 \hbar A}{b^2}.
\end{equation}

Note that the constant $A$ is given in terms of $\mu$ as follows $A=(1440 \pi^2)^{-1}[(1-4\mu)^{-4}-1]$ \cite{Hiscock1}, where for GUT strings one can take $\mu \simeq 10^{-6}$.

\subsection{Geodesic Equations}

Here we will study the geodesic motion and calculate the deflection angle in this framework. To describe geodesic motion in such a spacetime we recall the variational principle stated, which is written as
\begin{equation}
\delta \int \mathcal{L} \,\mathrm{d}s=0.
\end{equation}
We can obtain geodesic equations using the Lagrangian equation for the string metric \eqref{metric1}, as follows
\begin{widetext}
\begin{eqnarray}\label{lag1}
2\mathcal{L}=-\left(1-\frac{4 \pi \hbar A}{r^2}   \right)\dot{t}^2+\dot{r}^2+(1-4 \mu)^2 r^2  \left(1+\frac{16 \pi \hbar A}{r^2} \right)\dot{\varphi}^2.
\end{eqnarray}
\end{widetext}
We also characterized $\mathcal{L}$ by $+1, 0,$ and $-1$, for timelike, null, and spacelike geodesics, respectively.
To make further progress,  we consider the deflection of planar photons when $\theta =\pi/2$. Once again using the spacetime symmetries, we have the two constants of motion, namely $l$ and $\gamma$, and given as follows \cite{Boyer}
\begin{eqnarray}
p_{\varphi}&=&\frac{\partial \mathcal{L}}{\partial \dot{\varphi}}= (1-4 \mu)^2 r^2  \left(1+\frac{16 \pi \hbar A}{r^2}   \right)\dot{\varphi} =l,\\
p_{t}&=&-\frac{\partial \mathcal{L}}{\partial \dot{t}}= \left(1-\frac{4 \pi \hbar A}{r^2} \right) \dot{t}=\lambda.
\end{eqnarray}
Now, let us perform a coordinate transformation $r = 1/u(\varphi)$ by a new variable
$u(\varphi)$, which leads to the following identify
\begin{equation}\label{iden}
\frac{\dot{r}}{\dot{\varphi}}=\frac{\mathrm{d}r}{\mathrm{d}\varphi}=-\frac{1}{u^2}\frac{\mathrm{d}u}{\mathrm{d}\varphi}.
\end{equation}

According to Ref. \cite{Boyer}, without loss of generality we may set $\lambda=1$, and
using the condition $ u_{max} = 1/r_{min} = 1/b$ \cite{Iorio} to count the angle $\varphi$ from the point of closest approach. This allow us to
obtain the second constant of motion as 
\begin{equation}\label{l1}
l = (1- 4 \mu) b,
\end{equation}
and it is interesting to note that if $\mu \to 0$, then we have $l = b$. Finally using the Eqs. \eqref{lag1}, \eqref{iden} and \eqref{l1}, we get the following differential equation in terms of $u(\varphi)$ as follows
\begin{eqnarray}\label{diff1}
\frac{1}{u^4}\left( \frac{\mathrm{d}u}{\mathrm{d}\varphi} \right)^2+\frac{4 \pi A \hbar \zeta^2}{u^2 b^2 \eta^2}-\frac{(1-4 \mu)^2 \zeta^2}{u^4 b^2 \eta^2}+16 \pi (1-4\mu)^2 \hbar A+\frac{(1-4\mu)^2}{u^2}=0,
\end{eqnarray}
where for notational convention we use
\begin{eqnarray}
\zeta &=& 16 \pi A \hbar + \frac{1}{u^2},\\
\eta & = & 4 \pi A \hbar - \frac{1}{u^2}.
\end{eqnarray}

It is a straightforward calculation to find the equation for the light ray given by Eq. (20) from the differential equation  \eqref{diff1}. Let us simply neglect the quantum effects by setting $\hbar=0$ and differentiate Eq. \eqref{diff1} we find 
\begin{equation}
\frac{\mathrm{d}^2}{\mathrm{d}\varphi^2}u(\varphi)+(1-4\mu)^2 u(\varphi)=0.
\end{equation}

The solution of the above equation reads
\begin{equation}
u(\varphi)=C_1 \sin\left[(1-4\mu)\varphi\right]+C_2 \cos\left[(1-4\mu) \varphi\right].
\end{equation}

Now use the conditions $u(\varphi=0)=0$ and $u(\varphi=\pi/2)=1/b$, to find
\begin{equation}
u(\varphi)=\frac{\sin\left[(1-4\mu) \varphi\right]}{b}\frac{1}{\sin\left(\frac{(1-4\mu) \pi}{2}    \right)}.
\end{equation}

Finally we can write
\begin{equation}
u(\varphi)\simeq \frac{\sin\left((1-4\mu) \varphi\right)}{b},
\end{equation}
if we use $\sin\left(\frac{(1-4\mu) \pi}{2}\right)\simeq 1 $. Thus, it follows the light ray equation (20) given by
\begin{equation}
r(\varphi)=\frac{1}{u(\varphi)}=\frac{b}{\sin\left[\left(1-4\mu\right) \varphi\right]}.
\end{equation}

To find out the deflection angle we apply the perturbation method for solving the differential equation \eqref{diff1} and the solution of this differential equation in leading order terms can be expressed as (see Ref.  \cite{Boyer,weinberg})
\begin{equation}
\Delta \varphi =\pi+\hat{\alpha},
\end{equation}
where $\hat{\alpha}$ is the deflection angle to be calculated. 
Its solution in terms of the angle of deflection of light 
according to Ref. \cite{weinberg}, can be calculated as
\begin{equation}
\hat{\alpha}=2|\varphi_{u=1/b}-\varphi_{u=0}|-\pi,
\end{equation}
where 
\begin{equation}
\varphi=\int_0^{1/b} B ~\mathrm{d}u.
\end{equation}
with
\begin{align*}
B = \frac{b \left( 16 \mu^2+4\mu+1  \right) \left(8Ab^2\hbar u^4 \pi-18 A \pi \hbar u^2-u^2 b^2+1   \right)}{\sqrt{1-u^2 b^2}(1-u^2 b^2)}
\end{align*}
Amazingly, the deflection angle in the weak deflection limit approximation is found to be the exact result found by GBT \eqref{a1} given by
\begin{equation}\label{dif00}
\hat{\alpha}\simeq 4 \pi \mu+ 16 \pi \mu^2+\frac{6 \pi^2 \hbar A}{b^2}+\frac{24 \pi^2 \hbar \mu A}{b^2}+\frac{96 \pi^2 \mu^2 \hbar A}{b^2}.
\end{equation}

Thus, we found an agreement up to the second-order contribution terms in the cosmic string parameter $\mu$. The last term can be viewed as a third-order term if we consider quantum effects, hence overall our result includes contributions up to the third-order terms in the deflection angle. 

\section{A Magnetic Flux Cosmic String} 

We can extend our result by considering a more general scenario, namely we can consider a cosmic string carrying a magnetic flux $\Phi$ which interacts with some scalar field. In that case, the exterior metric of the magnetic flux cosmic
string with semiclassical gravitational effects was found by Guimaraes \cite{guimares}

\begin{eqnarray}\notag
\mathrm{d}s^2 &=& \left[1-\frac{4 \pi  \hbar }{\rho^2}\left(A(\gamma)-\frac{1}{2}B(\gamma) \right)\right]\left(-\mathrm{d}t^2+dz^2\right)+\mathrm{d}\rho^2+\left(1-4\mu\right)^2 \rho^2 \\
& \times & \left[1+\frac{16 \pi \hbar}{\rho^2} \left(A(\gamma)+\frac{1}{4}B(\gamma) \right)\right]\mathrm{d}\varphi^2.
\end{eqnarray}

The effect of magnetic flux is encoded in the the following two dimensionless quantities
$A (\gamma)$ and $B (\gamma)$, given by \cite{guimares}
\begin{eqnarray}
A (\gamma) &=& \omega_4 (\gamma)-\frac{1}{3}\omega_2 (\gamma) ,\\
B (\gamma) &=& 4 \left(\xi-\frac{1}{6}\right) \omega_2(\gamma).
\end{eqnarray}

In the present paper we consider the minimal  coupling case with $\xi=0$, where $\omega_2(\gamma)$ and $\omega_4(\gamma)$ are given as follows
\begin{eqnarray}\notag
\omega_2(\gamma) &=& - \frac{1}{8 \pi^2}\left\lbrace \frac{1}{3}-\frac{1}{2 \beta^2}\left[4 \left(  \gamma-\frac{1}{2}\right)^2-\frac{1}{3}    \right]    \right\rbrace ,\\\notag
\omega_4 (\gamma) &=& - \frac{1}{720 \pi^2}\left\lbrace 11-\frac{15}{\beta^2}\left[4 \left(\gamma-\frac{1}{2}    \right)^2-\frac{1}{3}      \right]+\frac{15}{8 \beta^4}\left[ 16 (\gamma-\frac{1}{2})^4-8(\gamma-\frac{1}{2})^2+\frac{7}{15}  \right]  \right\rbrace .
\end{eqnarray}

Furthermore $\gamma$ is called the fractional part of ${\Phi/\Phi_0}$ and lies in the interval $0 \leq \gamma <1$, $\beta=1-4\mu$, and $\Phi_0$ is the quantum flux given by $\Phi_0=2 \pi \hbar / e$ \cite{guimares,dowker,frolov,guimares2}. In the special case $\gamma \to 0$ corresponds to the absence of a magnetic flux. Further, as in Section II, it is very convenient to write the above metric in first in spherical coordinates and then  by taking $\theta=\pi/2$, we find
\begin{equation}
\mathrm{d}s^2=-\left[1-\frac{4 \pi  \hbar }{r^2}\left(A(\gamma)-\frac{1}{2}B(\gamma) \right)\right]\mathrm{d}t^2+\mathrm{d}r^2+(1-4\mu)^2 r^2 \left[1+\frac{16 \pi  \hbar}{r^2}\left(A(\gamma)+\frac{1}{4}B(\gamma)\right)\right]\mathrm{d}\varphi^2
\end{equation}

We can calculate the Gaussian optical curvature and see that the effect only the constant $A(\gamma)$ contributes as a linear term in $\hbar$. In order to see the effect of $B(\gamma)$ we need to include higher order terms in $\hbar$. In particular one can write the Gaussian optical curvature as  
\begin{equation}
K \simeq K_0+K_{correc.} 
\end{equation}
where 
\begin{equation}
K_0=-\frac{24 \pi A(\gamma) \hbar}{r^4}, 
\end{equation}
and 
\begin{equation}
 K_{corre} \propto \frac{B^2(\gamma) \hbar^2}{r^6} +\frac{A(\gamma)B(\gamma) \hbar^2}{r^6}+\frac{A^2(\gamma) \hbar^2}{r^6}
\end{equation}

In a similar way we can write the total deflection angle after we carry out the integration in the form
\begin{eqnarray}
\hat{\alpha}&=& \hat{\alpha}_0+\hat{\alpha}_{corre.}
\end{eqnarray}
where 
\begin{equation}\label{def11}
\hat{\alpha}_0\simeq 4 \pi \mu+ 16 \pi \mu^2+\frac{6 \pi^2 \hbar A(\gamma)}{b^2}+\frac{24 \pi^2 \hbar \mu A(\gamma)}{b^2}+\frac{96 \pi^2 \mu^2 \hbar A(\gamma)}{b^2},
\end{equation}
and 
\begin{equation}
\hat{\alpha}_{corre.} \propto \frac{B^2(\gamma) \hbar^2}{b^4}+ \frac{A^2(\gamma) \hbar^2}{b^4}+\frac{A(\gamma)B(\gamma) \hbar^2}{b^4}.
\end{equation}

This results shows that the deflection angle is affected by the magnetic flux of the cosmic string through the quantities  $A(\gamma)$ and $B(\gamma)$. We also see that, in the absence of a magnetic flux i.e., $\gamma \to 0$, as a limiting case of the deflection angle \eqref{def11}, we recover our result \eqref{dif00} found in Section II.  In this context, it is also important to note that, the effects of the second constant $B^2(\gamma)$ are negligible compared to $A(\gamma)$. In particular these terms are proportional to $\hbar^2$ which requires an agreement up to the fourth order terms in the deflection angle in order to fix the constants of proportionality.  However, as we already stated, in the present paper we only consider the terms linear in the $\hbar$.

\section{Semiclassical gravitational effects by a global monopole}

We now consider the semiclassical gravitational effects around the point-like global monopoles spacetime metric was found by Hiscock \cite{Hiscock2}, as follows 

\begin{equation}\label{metric2}
\mathrm{d}s^2=-\left[1+\frac{4 \pi \hbar (C+2D) }{r^2}\right]\mathrm{d}t^2+\left(1-\frac{8 \hbar \pi D}{r^2}\right)\mathrm{d}r^2+r^2 \alpha^2 \left(\mathrm{d}\theta^2+\sin^2\varphi \,\mathrm{d}\varphi^2 \right),
\end{equation}
where $C$ and $D$ stands for dimensionless constant with $\alpha^2=1-8 \pi \eta^2$, in which $\eta$ is known as the scale of gauge-symmetry breaking with $\eta =10^{16}$ GeV. Solving for null geodesics with $\mathrm{d}s^2 =0$, and considering the problem in the equatorial plane we find the global monopole optical metric is given by
\begin{equation}\label{dt2}
\mathrm{d}t^{2}=\frac{1-\frac{8 \hbar \pi D}{r^2}}{1+\frac{4 \pi \hbar (A+2B) }{r^2}}\mathrm{d}r^2+\frac{r^2 \alpha^2}{1+\frac{4 \pi \hbar (A+2B) }{r^2}}\mathrm{d}\varphi^2 \equiv {\mathrm{d}r^{\star}}^2+f(r^{\star})^2 \mathrm{d}\varphi^2,
\end{equation}
where we have used Regge-Wheeler tortoise type coordinate $r^{\star }$, such that
\begin{eqnarray}
\mathrm{d}r^{\star }&=&\left(\frac{1-\frac{8 \hbar \pi D}{r^2}}{1+\frac{4 \pi \hbar (C+2D) }{r^2}}\right)^{1/2}\mathrm{d}r, \\
f(r^{\star })&=&\frac{r \alpha}{\sqrt{1+\frac{4 \pi \hbar (C+2D) }{r^2}}}.
\end{eqnarray}

It is now straightforward to compute the Gaussian optical curvature $K$, which can be calculated by the following equation \cite{gibbons1}, which yields
\begin{equation}
K=\frac{\pi \hbar \left[D\hbar^2(C+2D)^2 \pi^2+\frac{D\hbar r^2 \pi (C+2D)}{8}+\frac{r^4 (C+3D)}{32}    \right]}{r^2\left[\pi \hbar (C+2D)+\frac{r^2}{4}\right]\left(\pi \hbar D-\frac{r^2}{8}\right)^2}.
\end{equation}

We also interested in the weak field limit, we can approximate the optical Gaussian curvature as 
\begin{equation}\label{GaussianCurvature2}
K\simeq  \frac{8 \pi \hbar }{r^4} \left(C+3D \right).
\end{equation}

Later on, we shall use this important result in the GBT for computing the angle of deflection of light.

\subsection{Deflection angle}

We can find the deflection angle by going through exactly the same procedure as in the cosmic string case. Calculating the 
the geodesic curvature to the curve $\gamma_R$, we find
\begin{eqnarray}\notag
\lim_{R\rightarrow \infty }\kappa (\gamma_{R}) &=&\lim_{R\rightarrow \infty
}\left\vert \nabla _{\dot{\gamma}_{R}}\dot{\gamma}_{R}\right\vert , \notag \\
&\rightarrow &\frac{1}{R}. 
\end{eqnarray}%

On the other hand, for very large radial distance Eq. \eqref{dt2} yields
\begin{eqnarray}\notag
\lim_{R\rightarrow \infty } \mathrm{d}t &=&\lim_{R\rightarrow \infty
}\left( \frac{R \sqrt{1-8 \pi \eta^2}}{\sqrt{1+\frac{4 \pi \hbar (C+2D) }{R^2}}}\right) \mathrm{d}\varphi\\
&\to & \sqrt{1-8 \pi \eta^2} R \, \mathrm{d}\varphi.  
\end{eqnarray}%

If we combine the last two equations we  find $
\kappa (\gamma_{R})\mathrm{d}t=\sqrt{1-8 \pi \eta^2} \mathrm{d}\,\varphi
$. In other words our optical geometry is not asymptotically  Euclidean. From GBT we have
\begin{equation}\label{GBT3}
\iint\limits_{\mathcal{S}_{R}}K\,\mathrm{d}A+\oint\limits_{\gamma_{R}}\kappa \,%
\mathrm{d}t\overset{{R\rightarrow \infty }}{=}\iint\limits_{\mathcal{S}%
_{\infty }}K\,\mathrm{d}A+ \sqrt{1-8 \pi \eta^2} \int\limits_{0}^{\pi + \hat{\alpha}}\mathrm{d}\varphi
=\pi.
\end{equation}

The key point relies on the fact that due to the conical topology of the spacetime, in the monopole case, we may choose the light ray as follows
\begin{equation}
r=\frac{b}{\sin\Big[\sqrt{1-8 \pi \eta^2 } \varphi\big]}.
\end{equation} 
 
If we substitute this relation into Eq. \eqref{GBT3} and we solve for the deflection angle then after we use the Gaussian curvature \eqref{GaussianCurvature2} we end up with the following integral
\begin{equation}
\hat{\alpha}=\pi \left[ \frac{1}{\sqrt{1-8 \pi \eta^2}}-1 \right]-\frac{1}{\sqrt{1-8\pi \eta^2}}\int\limits_{0}^{\pi}\int\limits_{\frac{b}{\sin\left(\sqrt{1-8 \pi \eta^2 } \varphi\right)}}^{\infty}\left[\frac{8 \pi \hbar \left(C+3D \right)}{r^4} \right]\sqrt{\det g^{op}}\mathrm{d}r^{\star}\mathrm{d}\varphi.
\end{equation}

The first term can be easily evaluated up to the second order in $\eta$, to find
\begin{equation}
\mathcal{I}_1=\pi \left[ \frac{1}{\sqrt{1-8 \pi \eta^2}}-1 \right]=4\pi^2 \eta^2+\mathcal{O}(\eta^4).
\end{equation}

Solving the integral we find for the second term the following result
\begin{eqnarray}\notag
\mathcal{I}_2 &=& -\frac{1}{\sqrt{1-8\pi \eta^2}}\int\limits_{0}^{\pi}\int\limits_{\frac{b}{\sin\left(\sqrt{1-8 \pi \eta^2 } \varphi\right)}}^{\infty}\left[\frac{8 \pi \hbar \left(C+3D \right)}{r^4} \right]\sqrt{\det g^{op}}\mathrm{d}r^{\star}\mathrm{d}\varphi \\
&=& -\frac{2 \pi^2 \hbar }{b^2}\left( C+3D \right)-\frac{8 \pi^3 \hbar \eta^2 }{b^2}\left(C+3D\right).
\end{eqnarray}

Finally putting together these results we find the total deflection angle
\begin{equation}\label{a2}
\hat{\alpha}\simeq 4 \pi^2 \eta^2-\frac{2 \pi^2 \hbar }{b^2}\left( C+3D \right)-\frac{8 \pi^3 \hbar \eta^2 }{b^2}\left(C+3D\right).
\end{equation}

\subsection{Geodesic Equations}

Applying the variational principle to the metric \eqref{metric2} we find the Lagrangian
\begin{eqnarray}\label{lag2}
2\mathcal{L}=-\left( 1+\frac{4 \pi \hbar (C+2D)}{r^2} \right)\dot{t}^2+\left( 1-\frac{8 \pi \hbar C}{r^2} \right) \dot{r}^2+\alpha^2 r^2 \dot{\varphi}^2.
\end{eqnarray}

The spacetime symmetries implies two constants of motion, namely $l$ and $\lambda$, given as follows \cite{Boyer}
\begin{eqnarray}
p_{\varphi}&=&\frac{\partial \mathcal{L}}{\partial \dot{\varphi}}= \alpha^2 r^2 \dot{\varphi}=l,\\
p_{t}&=&-\frac{\partial \mathcal{L}}{\partial \dot{t}}=\left( 1+\frac{4 \pi \hbar (C+2D)}{r^2}  \right)\dot{t}=\lambda.
\end{eqnarray}

Let us now introduce a new variable $u(\varphi)$, which is related to our old radial coordinate as follows $r=1/u(\varphi)$ and hence the following identity
\begin{equation}\label{iden1}
\frac{\dot{r}}{\dot{\varphi}}=\frac{\mathrm{d}r}{\mathrm{d}\varphi}=-\frac{1}{u^2}\frac{\mathrm{d}u}{\mathrm{d}\varphi}.
\end{equation}

Without loss of generality,  one can choose the second constant of motion as follows
\begin{equation}\label{l2}
l=\alpha \,b \,\sqrt{1-\frac{4 \pi C \hbar}{b^2}-\frac{8\pi D \hbar}{b^2}}
\end{equation}

We see from the last two equations that if we take the limit $\eta\to 0$ and $\hbar \to 0$,then $l=b$. Finally using Eqs. \eqref{lag2}, \eqref{iden1} and \eqref{l2}, in terms of $u(\varphi)$ we find the following equation
ting $\lambda=1$  and $u_{max}=1/r_{min}=1/b$  in leading order terms. In this ca
\begin{eqnarray}\label{diff2}
\frac{1}{u^4}\left( \frac{\mathrm{d}u}{\mathrm{d}\varphi} \right)^2-\frac{8 \pi D \hbar}{u^2}\left( \frac{\mathrm{d}u}{\mathrm{d}\varphi} \right)^2-\frac{4 \pi C \hbar \alpha^2}{u^6 b^2 \Xi \,\Theta^2}-\frac{8 \pi D \hbar \alpha^2}{u^6 b^2 \Xi \,\Theta^2}-\frac{\alpha^2}{u^8 b^2 \Xi \Theta^2}+\frac{\alpha^2}{u^2}=0,
\end{eqnarray}
where
\begin{eqnarray}
\Xi &=& -\frac{4 \pi C \hbar}{b^2}-\frac{8 \pi D \hbar }{b^2}+1, \\
\Theta & =& 4 \pi C \hbar +8 \pi D \hbar+\frac{1}{u^2}.
\end{eqnarray}

Let us now show that from \eqref{diff2} one can find the light ray equation (62) as well. To do so, we set $\hbar=0$ and differentiate Eq. \eqref{diff2} yielding 
\begin{equation}
\frac{\mathrm{d}^2}{\mathrm{d}\varphi^2}u(\varphi)+(1-8 \pi \eta^2) u(\varphi)=0.
\end{equation}

Which has the following solution
\begin{equation}
u(\varphi)=C_1 \sin\left[\sqrt{1-8 \pi \eta^2}\varphi\right]+C_2 \cos\left[\sqrt{1-8 \pi \eta^2}\varphi\right].
\end{equation}

Under the conditions $u(\varphi=0)=0$ and $u(\varphi=\pi/2)=1/b$, yields
\begin{equation}
u(\varphi)=\frac{\sin\left[\sqrt{1-8 \pi \eta^2} \varphi\right]}{b}\frac{1}{\sin\left(\frac{\sqrt{1-8 \pi \eta^2} \pi}{2}    \right)}.
\end{equation}

Hence we find
\begin{equation}
u(\varphi)\simeq \frac{\sin\left(\sqrt{1-8 \pi \eta^2} \varphi\right)}{b},
\end{equation}
where we have used $\sin\left(\frac{\sqrt{1-8 \pi \eta^2}\pi}{2}\right)\simeq 1 $. This equation is nothing else but our Eq. (62) given by
\begin{equation}
r(\varphi)=\frac{b}{\sin\left[\sqrt{1-8 \pi \eta^2}\varphi\right]}.
\end{equation}

It is well known that the solution of the differential Eq. \eqref{diff2} is given in terms of the familiar form  \cite{Boyer,weinberg}
\begin{equation}
\Delta \varphi =\pi+\hat{\alpha},
\end{equation}
where $\hat{\alpha}$ is the deflection angle to be calculated. Following the same arguments given in Ref. \cite{weinberg}, the deflection angle can be calculated as
\begin{equation}
\hat{\alpha}=2|\varphi_{u=1/b}-\varphi_{u=0}|-\pi.
\end{equation}
where 
\begin{equation}
\varphi=\int_0^{1/b} A(u)\, \mathrm{d}u
\end{equation}

Note that in the last equation $A$ is calculated by considering Taylor expansion series around $\hbar$ and $\eta$, given by

\begin{eqnarray}
A(u)= -\frac{8 \left(\pi \eta^2+\frac{1}{4}   \right) \left[\hbar \pi \left(C+2D+2Db^2 u^2   \right)-\frac{b^2}{2}    \right]}{b \sqrt{1-b^2 u^2}}
\end{eqnarray}

Amazingly, the deflection angle in the weak deflection limit approximation is found to be the exact result found by GBT \eqref{a2} given by
\begin{equation}
\hat{\alpha}\simeq 4 \pi^2 \eta^2-\frac{2 \pi^2 \hbar }{b^2}\left( C+3D \right)-\frac{8 \pi^3 \hbar \eta^2 }{b^2}\left(C+3D\right).
\end{equation}

Hence, we found an agreement up to the second-order contribution terms in the global monopole parameter $\eta$. 

\section{Time delay}

Time delay is yet another classical test of general relativity. It is a consequence of the time difference due to the massive gravitational field when the light ray follows two different paths to reach the observer. Let us consider the spherically symmetric spacetime by letting $\theta=\pi/2$, given by
\begin{equation}
ds^2 = -F(r) \mathrm{d}t^2 + H(r) \mathrm{d}r^2 + G(r)\mathrm{d}\varphi^2.
\end{equation}

Following \cite{weinberg}, we modify the equations of motions due to the presence of cosmic string by simply $J^2\to (1-4\mu)^2 r_0^2/F(r_0)$, where $J$ is a constant. Then the time required for the light ray to go from $r_0$ to $r$ in the case of a cosmic string spacetime is given by
\begin{equation}
t(r,r_0)=\int_{r_0}^{r} \left[\frac{\frac{H(r)}{F(r)}}{1-\frac{F(r)}{F(r_0)}\left(\frac{(1-4\mu) r_0}{r} \right)^2}  \right]^{1/2} \mathrm{d}r.
\end{equation}

On the other hand the total time required for the light ray to go from one point $r_1$ to a second point $r_2$ is given by
\begin{equation}
t_{12}=t(r_1,r_0)+t(r_2,r_0),
\end{equation}
where $r_0$ is the distance of closest approach from the cosmic string (global monopole), respectively. In this way, in leading order approximation from the metric (5) we find
\begin{equation}
t(r,r_0)=\int_{r_0}^{r} \left(1-\frac{4 \pi A \hbar}{r^2}   \right)^{-1/2} \left(1-   \frac{\left(1-\frac{4 \pi A \hbar}{r^2}   \right)(1-4\mu)^2 r_0^2}{\left(1-\frac{4 \pi A \hbar}{r_0^2}   \right)r^2}\right)^{-1/2} \mathrm{d}r
\end{equation}
or approximated as
\begin{equation}
t(r,r_0)\simeq \int_{r_0}^{r}\left( 1-\frac{r_0^2}{r^2} \right)^{-1/2}\left(1+\frac{4 \mu}{1-\frac{r^2}{r_0^2}}+\frac{4 \pi A \hbar}{r^2} \right) \mathrm{d}r
\end{equation}

The delay in time is given by \cite{weinberg}
\begin{equation}
\Delta T=2\left[t(r_1,r_0)+t(r_2,r_0)-\sqrt{r_1^2-r_0^2}-\sqrt{r_2^2-r_0^2} \right]
\end{equation}

Solving the integral (89) and neglecting the quantum effects we find 
\begin{equation}
\Delta T\simeq \frac{8 \mu r_0^2 \left(\sqrt{r_1^2-r_0^2} +\sqrt{r_2^2-r_0^2}   \right)}{\sqrt{r_1^2-r_0^2}  \sqrt{r_2^2-r_0^2} }.
\end{equation}

We shall now focus on the delay time in the spacetime of a point-like global monopole. In the case of a global monopole one has to introduce $J^2\to (1-8 \pi \eta^2) r_0^2/F(r_0)$. Thus, the time required for the light ray to go from $r_0$ to $r$ can be given as follows
\begin{equation}
t(r,r_0)=\int_{r_0}^{r} \left[\frac{\frac{H(r)}{F(r)}}{1-\frac{F(r)}{F(r_0)}\left(\frac{\sqrt{1-8 \pi\eta^2}\, r_0}{r} \right)^2}  \right]^{1/2} \mathrm{d}r.
\end{equation}

From the metric (53) we find
\begin{equation}
t(r,r_0)\simeq \int_{r_0}^{r}\left( 1-\frac{r_0^2}{r^2} \right)^{-1/2}\left(1+\frac{4 \pi \eta^2 }{1-\frac{r^2}{r_0^2}}-\frac{2 \pi \hbar (C+2D)}{r^2} \right) \mathrm{d}r
\end{equation}

Solving the integral (93) and neglecting the quantum effects from Eq. (90) we find the following result
\begin{equation}
\Delta T\simeq \frac{8 \pi \eta^2 r_0^2 \left(\sqrt{r_1^2-r_0^2} +\sqrt{r_2^2-r_0^2}   \right)}{\sqrt{r_1^2-r_0^2}  \sqrt{r_2^2-r_0^2} }.
\end{equation}

\section{Conclusion }

In this text, we have calculated the quantum corrected deflection angle of light in the space-time of topological defects. In particular we have used the exterior quantum corrected metric by a cosmic string and a point-like global monopole found recently by Hiscock. We have applied the GBT to the optical geometry in these two cases and modified the integration domain by taking into account the global conical topology of the space-time. We found that the deflection angle is affected by the quantum effects in an interesting way. It is interesting to note that in the case of the cosmic string the deflection angle can be further extended if we introduce a magnetic flux cosmic string. More importantly, the modification of the integration domain 
led us to the correct result for the deflection angle up to the second order in $\mu$ and $\eta$, respectively. Finally, we have verified our results in terms of the standard geodesics approach and shown the agreement with GBT approach which confirms our achieved the desired results.   One can  observe that
the deflection of light is always positive and this indicates that light rays always bend towards cosmic strings and monopoles.
This results clearly suggest a link between the gravitational fields of topological defects 
and lensing features. On the other hand, our analysis of the time delay in the cosmic string/global monopole spacetime shows that the delay in time is proportional to the linear mass density of the  cosmic string and global monopole parameter. 
We conclude our study with the hope that the sharp  deflection angles  resulting from global monopole and cosmic string may provide us a way to probe the spacetime property of topological defects on the astronomical observations.  \\
\newpage
\textbf{Acknowledgement}: \\

FR   is grateful to the Inter-University Centre for Astronomy and Astrophysics
(IUCAA), India, for providing Associateship Programme.  Also FR is  thankful to DST, Government of India for providing financial support
under the SERB programme.  We are also thankful to the referee for his valuable and constructive suggestions.

\end{document}